\documentclass[english,twocolumn,showpacs,pra,amsmath,amssymb,english]{revtex4}
\usepackage[T1]{fontenc}
\usepackage[latin9]{inputenc}
\usepackage{amstext}
\usepackage{graphicx}

\makeatletter
\@ifundefined{textcolor}{}
{%
 \definecolor{BLACK}{gray}{0}
 \definecolor{WHITE}{gray}{1}
 \definecolor{RED}{rgb}{1,0,0}
 \definecolor{GREEN}{rgb}{0,1,0}
 \definecolor{BLUE}{rgb}{0,0,1}
 \definecolor{CYAN}{cmyk}{1,0,0,0}
 \definecolor{MAGENTA}{cmyk}{0,1,0,0}
 \definecolor{YELLOW}{cmyk}{0,0,1,0}
}

\usepackage[T1]{fontenc}
\usepackage[latin9]{inputenc}
\usepackage{amstext}
\usepackage{graphicx}

\makeatletter
\@ifundefined{textcolor}{}
{%
 \definecolor{BLACK}{gray}{0}
 \definecolor{WHITE}{gray}{1}
 \definecolor{RED}{rgb}{1,0,0}
 \definecolor{GREEN}{rgb}{0,1,0}
 \definecolor{BLUE}{rgb}{0,0,1}
 \definecolor{CYAN}{cmyk}{1,0,0,0}
 \definecolor{MAGENTA}{cmyk}{0,1,0,0}
 \definecolor{YELLOW}{cmyk}{0,0,1,0}
}

\makeatother

\usepackage{babel}

\makeatother

\usepackage{babel}
\begin{document}

\preprint{This line only printed with preprint option}

\title{Violation of Bell inequality in perfect translation invariant systems}

\author{Zhao-Yu Sun, Yu-Yin Wu, Hai-Lin Huang, Bo-Jun Chen}

\affiliation{School of Electrical and Electronic Engineering, Wuhan Polytechnic
University, Wuhan 430000, China.}

\author{Bo Wang}

\affiliation{ENN Group Co., Ltd., Langfang, Hebei 065001, China}
\begin{abstract}
Bell inequalities and nonlocality have been widely studied in one-dimensional
quantum systems. As a kind of quantum correlation, it is expected
that bipartite nonlocaity should be present in quantum systems, just
as bipartite entanglement does. Surprisingly, for various models,
two-qubit states do not violate Bell inequalities, i.e., they are
local. Recently, it is realized that the results are related to the
monogamy trade-off obeyed by bipartite Bell correlations, thus it
is believed that for general translation invariant systems, two-qubit
states should not violate the Bell inequality(Oliveira, EPL 100, 60004
(2012)). In this report, we will demonstrate that in perfect translation
invariant systems, the Bell inequality can be violated. A nontrivial
model is constructed to confirm the conclusion.
\end{abstract}
\maketitle

\section{introduction}

Quantum correlation has been a hot subject in the fields of quantum
information\cite{information} and condensed matter physics\cite{BOOK_QPT}
for many years. Quantum entanglement, which is the most famous feature
of quantum correlation, has attracted much attention in low-dimensional
quantum systems. The simplest form of entanglement is bipartite entanglement,
which describe the entanglement between two parts of a system. Bipartite
entanglement is widespread in quantum systems, for example, quantum
spin systems and Hubbard models. It's found that entanglement measures
show a singularity or a maximum in the vicinity of the quantum phase
transition (QPT) points in various models.\cite{QE_QPT} Suppose entanglement
in condensed matters can be measure experimentally, it can be used
as a valuable detector of phase transitions, just as traditional order
parameters do. It has been found that bipartite entanglement is short-ranged,
which results from the so-called monogamy trade-off obeyed by bipartite
entanglement.\cite{mono_entanglenet} Recently, the definition of
entanglement has been generalized to multipartite setting.\cite{Hierarchies_Multipartite_Entanglement}

Another closed related concept is quantum nonlocality, indicated by
the violation of Bell inequalities.\cite{Bell_Bell_Inequalityies}
Bell inequalities are experimentally testable, and many relevant experiments
have been carried out, from the first actual experiment by Freedman
and Clauser in 1972 to modern experiments nowadays.\cite{Bell_experiment,Bell_modern_experiment}
From this perspective, Bell inequalities and nonlocality greatly promote
the understanding of foundations of quantum mechanics. A Bell inequality
for two-party systems is derived by Clauser, Horne, Shimony and Holt.
It's usually called the CHSH inequality.\cite{CHSH} For two-qubit
states, an analytic formula for the violation of the CHSH inequality
has been found by Horodecki,\cite{Horodecki_BFV_twoQubitState} which
has been widely used to investigate bipartite nonlocality in quantum
spin models.

For a long time, non-locality and entanglement were regarded as equivalent
concepts. Then it is realized by Werner that for mixed states, an
entangled state may not violate any Bell inequality.\cite{Bell_mixedstate}
An even striking difference emerges when Bell inequality is applied
for two-qubit subsystems in low-dimensional spin models. As we have
mentioned, bipartite quantum correlation (such as entanglement) is
widely observed in quantum systems. However, it turns out that the
CHSH inequality is not violated for various one-dimensional quantum
models, including the transverse field Ising and XY model, the XY
model with multiple spin interactions, the XXZ model, and matrix product
models.(add topo QPT model)\cite{Bell_inequalitiesQPTs_XXZ_model,Correlation_nonlocalityQPTS_several_systems,BFV_Topological_QPT,Bell_ladder_MPS,Bell_QPT_models}

Recently, it is realized by Oliveira \cite{nonviolation} that the
nonviolation of biparite Bell inequality in one-dimensional system
is related to the monogamy trade-off obeyed by bipartite Bell correlations.\cite{Bell_monogon}
Thus, it was believed that two-qubit states should not violate the
Bell inequality for general translation invariant systems.\cite{nonviolation}
In this brief report, we will demonstrate that in perfect translation
invariant systems with even number of spins, the CHSH inequality can
be violated. As an example, we study the violation of Bell inequality
in a nontrivial perfect-translation-invariant model, which can be
projected to a DNA-style ladder model.

This paper is organized as follows. In Sec, II, we briefly review
the monogamy of bipartite nonlocality, and analyze the possible form
of bipartite nonlocality in translation invariant systems. In Sec.
III, we construct a quantum model which violates the bipartite CHSH
inequality. A summary is given in Sec. IV.

\section{bell-chsh inequality and monogamy}

For a two-qubit state $\hat{\rho}$, one first defines a matrix $\hat{M}$
as 
\[
M_{ij}(\hat{\rho})=\textrm{Tr}[\hat{\rho}\cdot\hat{\sigma}_{i}\otimes\hat{\sigma}_{j}],
\]
where $\hat{\sigma}_{i}$ are the three Pauli matrices, i.e., $\hat{\sigma}_{1,2,3}=\hat{\sigma}_{x,y,z}$.
Then one constructs a symmetric matrix $\hat{M}\ensuremath{^{{\rm T}}}\hat{M}$,
and finds its two largest eigenvalues $\lambda_{1}$ and $\lambda_{2}$.
According to CHSH inequality and Horodecki's formula, for any state
described by a realistic local theory, it should hold that\cite{Horodecki_BFV_twoQubitState}
\begin{equation}
\mathcal{B}(\hat{\rho})=2\sqrt{\lambda_{1}+\lambda_{2}}\le2.\label{eq:Horodecki}
\end{equation}
$\mathcal{B}$ is called a violation measure. For some state $\hat{\rho}$,
if the above inequality is violated, the state cannot be characterized
by any realistic local theory, in other words, $\hat{\rho}$ is non-local.
The Bell inequality is maximally violated by the Bell states, for
which the violation measure is $2\sqrt{2}$.

Monogamy of nonlocality describes the constraints for the distribution
of nonlocality between parties in the system. In an $N$-spin composite
system($N\ge3$), for arbitrary three sites $i$, $j$ and $k$, it
is found that\cite{Bell_monogon}

\[
\mathcal{B}^{2}(\hat{\rho}_{ij})+\mathcal{B}^{2}(\hat{\rho}_{jk})\le8.
\]

From the monigamy ineqality one can check that, if $\hat{\rho}_{ij}$
violates the Bell inequality, $\hat{\rho}_{jk}$ should not violate
the inequality, and vice versa. It indicates that the spin $j$ can
share nonlocal correlation only with spin $i$ or with spin $k$,
but never both.

\begin{figure}
\includegraphics{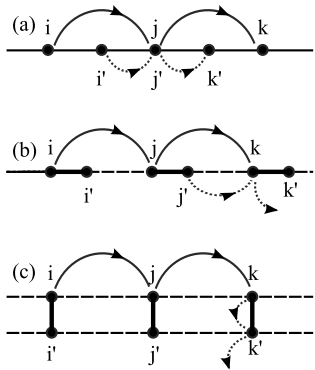}\caption{\label{fig:1D}Several typical one-dimensional topologies. (a) The
simplest topology with perfect translational invariance. (b) A periodic
dimer chain. (c) A periodic ladder.}
\end{figure}

The monigamy of nonlocality has been used to explain the nonviolation
of Bell inequality in quantum spin systems. For two spins $i$ and
$j$ in a system, if one can find a third spin $k$ such that 
\begin{equation}
\hat{\rho}_{ij}=\hat{\rho}_{jk}\label{eq:equal_rho}
\end{equation}
the monigamy inequaltity will reduce to $2\mathcal{B}^{2}(\hat{\rho}_{ij})\le8$,
i.e., $\mathcal{B}(\hat{\rho}_{ij})\le2$. In translation invariant
models, the third spin is usually easy to identify, as shown in Fig.
\ref{fig:1D}(a). 

For some models with slightly complex topology, the physical picture
is still clear. A dimerized spin model is shown in Fig. \ref{fig:1D}(b),
which consists of two sublattices. For the spins on the same sublattice,
for example, spins $i$ and $j$, the monigamy applies, thus the Bell
inequality should not be violated. However, the monigamy inequaltity
has no constrain on two nearest-neighboring spins on different sublattices,
such as spins $j'$ and $k$, since we cannot find a third spin satisfying
Eq. (\ref{eq:equal_rho}). As a result, the Bell inequality can be
violated for two nearest neighbors in the dimerized model.\cite{nonviolation}
Similar result has been reported in a spin ladder model, whose ground
state can be expressed as the matrix product states.\cite{Bell_ladder_MPS}
In a ladder lattice (see Fig. \ref{fig:1D}(c)), the monigamy inequality
suggests that the two spins on a leg should not violate the Bell inequality.
However, the monigamy inequaltity has no constrain on the two spins
on a rung. In fact, the state of the rung indeed violates the Bell
inequality. In these two situations, bipartite nonlocality is observed
for two nearest-neighboring sites which reside on two different sublattices
of the systems.

\begin{figure}
\includegraphics{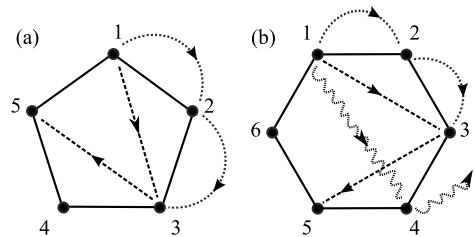}\caption{\label{fig:polygon} Topology for (a) a regular pentagon and (b) a
regular hexagon.}
\end{figure}

In this paper, we report that the Bell inequality can even be violated
in perfect translation invariant models, i.e., all the $m$ spins
locate at the vertices of a regular $m$-sided polygon. See Fig. \ref{fig:polygon}.
Firstly, we consider a regular polygon with odd number of spins, shown
in Fig. \ref{fig:polygon}(a). For any two sites on the polygon, such
as nearest neighbors (spins $1$ and $2$) and next-nearest neighbors
(spins $1$ and $3$), one can always find the third spin satisfying
Eq. (\ref{eq:equal_rho}). Thus, Bell inequality should never be violated
in perfect translation invariant models with odd number of spins.

Next, we discuss models with even number of spins. In Fig. \ref{fig:polygon}(b)
we illustrate a regular polygon with $m=6$. For nearest neighbors
and next-nearest neighbors, we can easily identify the third spin
which satisfies Eq. (\ref{eq:equal_rho}). However, for spins $1$
and $4$, no third spin satisfies Eq. (\ref{eq:equal_rho}). Thereby,
for any regular polygon with even number of spins, only the spin pairs,
which locate at the two ends of the longest diagonals of the polygon,
are not subject to the constraints of the monogamy inequality. They
can violate the Bell inequality, in other words, they can be nonlocal.

Consequently, bipartite nonlocality can be observed in perfect translation
invariant models under two necessary but not sufficient conditions.
First, the models must have even number of spins. Second, bipartite
nonlocality can only be present for spin pairs at the two ends of
the longest diagonals of the polygon.

\section{model}

\begin{figure}
\includegraphics{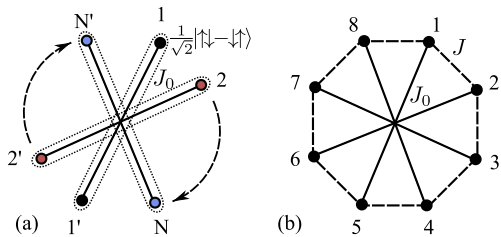}\caption{\label{fig:our_model} Using dimers to construct a regular polygon.}
\end{figure}

In this section, we will construct a concrete model to illustrate
the violation of Bell inequality. Let's consider a dimer consisting
of two spins $1$ and $1'$ which interact with each other via an
anti-ferromagnetic Heisenberg exchange $J_{0}(S_{1}^{x}S_{2}^{x}+S_{1}^{y}S_{2}^{y}+\Delta_{0}S_{1}^{z}S_{2}^{z})$,
with the coupling constant $J_{0}>0$ and the anisotropic parameter
$\Delta_{0}$. We need to set $\Delta_{0}>-1$, so that the ground
state is $|\Psi^{-}\rangle=\frac{1}{\sqrt{2}}(|\uparrow\downarrow\rangle-|\downarrow\uparrow\rangle)$.
In the filed of condensed matter physics, $|\Psi^{-}\rangle$ is usually
called a singlet. While in quantum information theory, it is one of
the four Bell states which violate the Bell inequality maximally.
This Bell state is just the starting point to construct our model.
Similarly to the dimer $(1,1^{'})$, we set up other dimers $(2,2^{'})$,
$(3,3^{'})$, ... $(N,N')$, as shown in Fig. \ref{fig:our_model}(a).
All these spins occupy the vertices of a regular $2N$-sided polygon,
and every dimer $(i,i')$ is a longest diagonal of the polygon.

Then, along the sides of the polygon, let the nearest-neighboring
spins have a Heisenberg exchange path, with the coupling constant
$J$ and $\Delta$(see Fig. \ref{fig:our_model}(b)). Finally, we
obtain a quantum system described by the following Hamiltonian

\[
\begin{array}{cl}
\hat{H}= & J_{0}\sum_{\{i,i'\}}(S_{i}^{x}S_{i'}^{x}+S_{i}^{y}S_{i'}^{y}+\Delta_{0}S_{i}^{z}S_{i'}^{z})\\
 & +J\sum_{\{i,j\}}(S_{i}^{x}S_{j}^{x}+S_{i}^{y}S_{j}^{y}+\Delta S_{i}^{z}S_{j}^{z}),
\end{array}
\]
where the first term summarizes over all the interactions along the
diagonals of the polygon with $i'=i+N$, and the second term summarizes
over all interactions along the sides with $j=i+1$. The model has
even number ($2N$) of spins and is perfectly translation invariant.

Qualitative analysis is adequate to confirm the existence of bipartite
nonlocality in the system. In the case $\frac{|J|}{J_{0}}\ll1$, the
system can be regarded as an ensemble of $N$ uncoupled dimers, illustrated
in Fig. \ref{fig:our_model}(a). The ground states of these dimers
are just the Bell state $|\Psi^{-}\rangle$, thus the Bell inequality
is violated maximally. The violation will retain in the thermodynamic
limit, since the ground state of the system can always be expressed
as the product of Bell states, regardless the size of the system.

\begin{figure}
\includegraphics[scale=0.68]{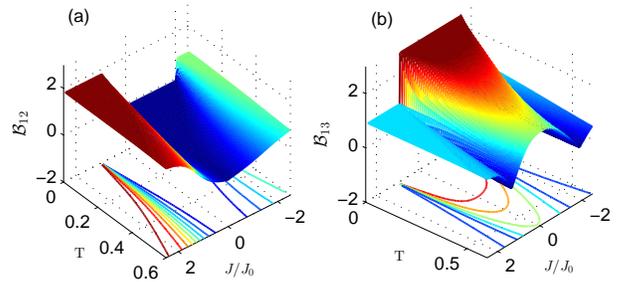}\caption{\label{fig:violation_N4} The violation measure as a function of temperature
$T$ and coupling constant $J_{0}/J$ for (a) two nearest-neighboring
spins and (b) two spins on the diagonal in a regular square model.}
\end{figure}

We use $J_{0}$ as the energy unit. In Fig. \ref{fig:violation_N4},
the violation measure $\mathcal{B}$ is studied as a function of $\frac{J}{J_{0}}$
and temperature $T$ for a regular square, with $\Delta=\Delta_{0}=1$.
According to the theory in Sec II, nonlocality can only be present
between two spins on a diagonal, i.e., spins $1$ and $3$. For nearest-neighboring
spins $1$ and $2$, as shown in Fig. \ref{fig:violation_N4}(a),
the Bell inequality is never violated. While for spins $1$ and $3$,
as shown in Fig. \ref{fig:violation_N4}(b), at zero temperature,
the Bell inequality is maximally violated ($\mathcal{B}_{13}=2\sqrt{2}$)
for $-2<\frac{J}{J_{0}}<1$. The violation region is far beyond the
strict region $\frac{|J|}{J_{0}}\ll1$. At finite temperatures, one
can see that the nonlocality is weakened by thermal fluctuation, and
the violated region of the Bell inequality is reduced as the increase
of the temperature. When the temperature is high enough, i.e., $T>0.43$,
the Bell inequality is not violated for any $J$. 

\begin{figure}
\includegraphics{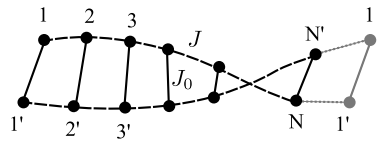}\caption{\label{fig:ladder}A distorted ladder with perfect translational invariance.}
\end{figure}

One may argue that the violation of Bell inequality in this model
is trivial, since in typical one-dimensional infinite systems long-range
interaction cannot be larger than short-range interaction. In fact,
we find that our model can be projected into a DNA-style ladder (Fig.
\ref{fig:ladder}). The diagonals of the polygon in Fig. \ref{fig:our_model}
are transformed to rungs of the ladder, and the sides of the polygon
forms the legs of the ladder. Special attention should be paid to
the boundary condition of the ladder. In the original polygon, spin
$N$ interacts with spin $1^{'}$, and spin $N^{'}$ interacts with
spin $1$. Thus the resulting ladder is neither an open ladder nor
a closed one. Instead, it is distorted, just like the structure of
DNA. Consequently, our model has a one-dimensional correspondence,
thus it is a nontrivial model to violate the Bell inequality.

\begin{figure}
\includegraphics{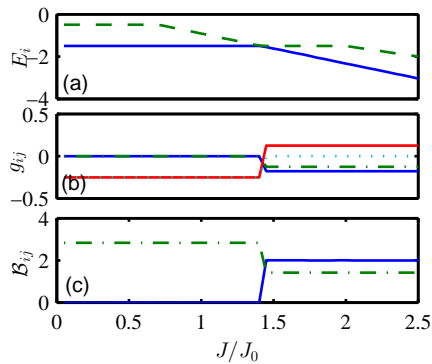}\caption{\label{fig:Bell_QPT} (Color Online) (a) Lowest-lying energy levels.
(b) Spin correlation functions. On the right part of the figure, the
four lines from top to bottom are corresponding to $\langle S_{1}^{x}S_{3}^{x}\rangle$,$\langle S_{1}^{z}S_{3}^{z}\rangle$,$\langle S_{1}^{z}S_{2}^{z}\rangle$,$\langle S_{1}^{x}S_{2}^{x}\rangle$,
respectively. (c) Violation measures $\mathcal{B}_{12}$(dash-dot
line) and $\mathcal{B}_{13}$(solid line). Parameters are set as $\Delta=0$,$\Delta_{0}=1$,$T=1.5\times10^{-5}$
and $N=4$.}
\end{figure}

In fact, just as the ladder model, our model has a rich phase diagram
in the thermodynamic limit. We can analyze the phase in several limits.
In the case $\frac{|J|}{J_{0}}\ll1$, the system is in a diagonal-singlet
phase. In the case $\frac{|J|}{J_{0}}\gg1$, the system reduces to
a normal spin-$\frac{1}{2}$ XXZ model. The XXZ model can be solved
exactly.\cite{solve_XXZ} The phase diagram contains a ferromagnetic
phase ($\Delta<-1$), a spin-fluid phase ($-1<\Delta<1$) and a Néel
phase($\Delta>1$). In fact, there can be other novel phases in intermediate
parameter regions. For its one-dimensinoal correspondence, i.e., the
ladder model, eight phases have been identified using various boundary
conditions.\cite{ladder} Thus various QPTs are expected in this model.
As for quantum entanglement, the ability of the violation measure
$\mathcal{B}$ to detect QPTs has been studied in many models, and
basic mechanism has been clarified.\cite{Bell_inequalitiesQPTs_XXZ_model,Bell_ladder_MPS,Bell_QPT_models,BFV_Topological_QPT,Correlation_nonlocalityQPTS_several_systems,nonviolation}
It has been realized that $\mathcal{B}$ can be expressed as a function
of the spin-spin correlation functions, in other words, the elements
of the density matrix $\hat{\rho}$. That's why the violation measure
can be used to detect QPTs. For the model considered in this paper,
for any two spins $i$ and $j$, the violation measure turns out to
be 
\begin{equation}
\mathcal{B}_{ij}=8\max\{\sqrt{\langle S_{i}^{x}S_{j}^{x}\rangle^{2}+\langle S_{i}^{z}S_{j}^{z}\rangle^{2}},\sqrt{2}\vert\langle S_{i}^{x}S_{j}^{x}\rangle\vert\}.\label{eq:Bij}
\end{equation}
In Fig. \ref{fig:Bell_QPT}, we have shown $\mathcal{B}_{ij}$, $\langle S_{i}^{x}S_{j}^{x}\rangle$,
$\langle S_{i}^{z}S_{j}^{z}\rangle$ and two lowest-lying energy levels
for a finite model. One can see that when an energy level crossing
occurs, $\langle S_{i}^{x}S_{j}^{x}\rangle$ and $\langle S_{i}^{z}S_{j}^{z}\rangle$
changes abruptly. Then according to Eq. (\ref{eq:Bij}), $\mathcal{B}_{ij}$
captures the sudden-change in $\langle S_{i}^{x}S_{j}^{x}\rangle$
and $\langle S_{i}^{z}S_{j}^{z}\rangle$, and consequently, captures
the change in the ground state. Our results just illustrate the ability
of $\mathcal{B}$ to capture the level crossing in a finite model.
Though QPTs occur in infinite systems, the intrinsic mechanism is
similar. In fact, it has been found that $\mathcal{B}$ can capture
QPTs for various models, no matter whether the Bell inequality is
violated or not. Since the ability of $\mathcal{B}$ in detecting
QPTs has nothing to do with the violation of Bell inequality, and
basic understanding has been achieved in previous papers, we will
not study this issue in this report.

\section{summaries}

In this report, we have re-considered the violation of Bell inequality
in translation invariant systems. The distribution of bipartite nonlocality
between different spins is constrained by the monogamy $\mathcal{B}^{2}(\hat{\rho}_{ij})+\mathcal{B}^{2}(\hat{\rho}_{jk})\le8$.
For two spins $i$ and $j$, suppose we can find a third spin such
that $\hat{\rho}_{ij}=\hat{\rho}_{jk}$, $\hat{\rho}_{ij}$ should
never violate the Bell inequality. As a result, for a perfect translation
invariant system with odd sites, Bell inequality should never be violated.
While for a system with even sites, the Bell inequality can be violated
for, and only for, the two spins at the two ends of the longest diagonals
of the polygon.

To illustrate our theory, we have constructed an artificial model
consisting of weakly coupled dimers, which contains even number of
spins and is perfect translation invariant. Qualitative analysis confirms
that the Bell inequality is indeed violated when the long-range interaction
(along the diagonals) is much larger than the short-range interaction
(along the sides) in the polygon. Furthermore, we find the model can
be projected into a ladder model with a DNA-style distortion. In several
limits, the properties of the phases are identified. Thereby, our
model has an one-dimensional correspondence with a very rich phase
diagram, and it is a nontrivial model to violate the Bell inequality.
\begin{acknowledgments}
The research was supported by the National Natural Science Foundation
of China (No. 11204223). This work was also supported by the Talent
Scientific Research Foundation of Wuhan Polytechnic University (Nos.
2012RZ09 and 2011RZ15).\end{acknowledgments}

\end{document}